%Paper: 9111049
%From: Norisuke Sakai <nsakai@cc.titech.ac.jp>
%Date: Mon, 25 Nov 91 11:33:23 +0900

\input phyzzx
%
% The definitions of \abstract, \ack, \refout and \figout
% in phyzzx.tex are changed to make "Abstract",
% "Acknowledgments", "References" and "Figure Captions"
% bold face and titlestyle.
%
\def\abstract{\vskip\frontpageskip
   \centerline{\bf {\titlestyle Abstract}} \vskip\headskip}
\def\ack{\par\penalty-100\medskip \spacecheck\sectionminspace
   \line{\hfil {\titlestyle {\bf Acknowledgements}} \hfil}
   \nobreak\vskip\headskip}
\def\refout{\par \penalty-400 \vskip\chapterskip
  \spacecheck\referenceminspace \immediate\closeout\referencewrite
  \referenceopenfalse
  \line{\hfil {\titlestyle {\bf References}} \hfil}\vskip\headskip
  \input \jobname.refs}
\def\figout{\par \penalty-400 \vskip\chapterskip
  \spacecheck\referenceminspace \immediate\closeout\figurewrite
  \figureopenfalse
  \line{\hfil {\titlestyle {\bf Figure Captions}} \hfil}
  \vskip\headskip
  \input \jobname.figs}
%
% The definition of \chapter is changed.
%
\def\chapter#1{\par \penalty-300 \vskip\chapterskip
   \spacecheck\chapterminspace
   \chapterreset \titlestyle{\S{\bf \chapterlabel \ \ #1}}
   \nobreak\vskip\headskip \penalty 30000
   \wlog{\string\chapter\ \chapterlabel}}
%
% Additional definitions to REFMACRO.TEX in phyzzx.tex:
%     \refmark and \refitem for Prog. Theor. Phys.
%
\def\Prefs{\let\therefmark=\refmark \let\therefitem=\Prefitem}
\def\refmark#1{\attach{\scriptscriptstyle #1)}}
\def\Prefitem#1{\refitem{#1)}}
\def\Prefmark#1{#1)}
\Prefs
%
% Examples
% ... matrix models \refmark{\bipz) - \amjukr}
% ... the Liouville theory\rlap.\refmark{\dika) - \fuinsu}
% ... Refs.\ \Prefmark{\dika) -- \david}
%
%%%%%%%%%%%%%%%%%%%%%%%%% References %%%%%%%%%%%%%%%%%%%%%%%%%%%%%%%%
%
\REFS\brka{E. Br\'ezin and V.A. Kazakov,
           {\it Phys.\ Lett.\ }{\bf 236B} (1990), 144;
           M. Douglas and S. Shenker,
           {\it Nucl.\ Phys.\ }{\bf B335} (1990), 635;
           D.J. Gross and A.A. Migdal,
           {\it Phys.\ Rev.\ Lett.\ }{\bf 64} (1990), 127;
           {\it Nucl.\ Phys.\ }{\bf B340} (1990), 333.}
\REFSCON\gmil{D.J. Gross and N. Miljkovi\'c,
           {\it Phys.\ Lett.\ }{\bf 238B} (1990), 217;
           E. Br\'ezin V.A. Kazakov and A. Zamolodchikov,
           {\it Nucl.\ Phys.\ }{\bf B338} (1990), 673;
           P. Ginsparg and J. Zinn-Justin,
           {\it Phys.\ Lett.\ }{\bf 240B} (1990), 333;
           G. Parisi, {\it Phys.\ Lett.\ }{\bf 238B} (1990), 209, 213;
           J. Ambj\o rn, J. Jurkiewicz and A. Krzywicki,
           {\it Phys.\ Lett.\ }{\bf 243B} (1990), 373;
           D.J. Gross and I.R. Klebanov,
           {\it Nucl.\ Phys.\ }{\bf B344} (1990), 475;
           {\it Nucl.\ Phys.\ }{\bf B354} (1991), 459.}
\REFSCON\dika{J. Distler and H. Kawai,
           {\it Nucl.\ Phys.\ }{\bf B321} (1989), 509;
           J. Distler, Z. Hlousek and H. Kawai,
           {\it Int.\ J. of Mod.\ Phys.\ }{\bf A5} (1990), 391; 1093;
           F. David, {\it Mod.\ Phys.\ Lett.\ }{\bf A3} (1989), 1651.}
\REFSCON\seirev{N. Seiberg,
           {\it Prog.\ Theor.\ Phys.\ Suppl.\ }{\bf 102} (1990), 319.}
\REFSCON\poltalk{J. Polchinski,
           in {\it Strings '90}, eds. R. Arnowitt et al.,
           (World Scientific, Singapore, 1991) p.\ 62;
           {\it Nucl.\ Phys.\ }{\bf B357} (1991), 241.}
\REFSCON\kitarev{Y. Kitazawa,
           Harvard preprint HUTP--91/A034 (1991).}
\REFSCON\gouli{M. Goulian and M. Li,
           {\it Phys.\ Rev.\ Lett.\ }{\bf 66} (1991), 2051;
           A. Gupta, S. Trivedi and M. Wise,
           {\it Nucl. Phys.\ }{\bf B340} (1990), 475.}
\REFSCON\dfku{P. Di Francesco and D. Kutasov,
          {\it Phys.\ Lett.\ }{\bf 261B} (1991), 385.}
\REFSCON\kita{Y. Kitazawa,
          {\it Phys.\ Lett.\ }{\bf 265B} (1991), 262.}
\REFSCON\sataco{N. Sakai and Y. Tanii,
           {\it Prog.\ Theor.\ Phys.\ }{\bf 86} (1991), 547.}
\REFSCON\dotsenko{V.S. Dotsenko,
            Paris preprint PAR--LPTHE 91--18 (1991).}
\REFSCON\taya{Y. Tanii and S. Yamaguchi,
            {\it Mod.\ Phys.\ Lett.\ }{\bf A6} (1991), 2271.}
\REFSCON\kostov{I. Kostov,
           {\it Phys.\ Lett.\ }{\bf 215B} (1988), 499; D.V. Boulatov,
           {\it Phys.\ Lett.\ }{\bf 237B} (1990), 202; S. Ben-Menahem,
           {\it Nucl. Phys.\ }{\bf B364} (1991), 681.}
\REFSCON\grklne{D.J. Gross, I.R. Klebanov and M.J. Newman,
           {\it Nucl.\ Phys.\ }{\bf B350} (1991), 621;
           D.J. Gross and I.R. Klebanov,
           {\it Nucl.\ Phys.\ }{\bf B352} (1991), 671;
           {\it Nucl.\ Phys.\ }{\bf B359} (1991), 3;
           G. Moore, Yale and Rutgers preprint YCTP--P8--91,
           RU--91--12 (1991);
           A.M. Sengupta and S.R. Wadia,
           {\it Int.\ J. of Mod.\ Phys.\ }{\bf A6} (1991), 1961;
           G. Mandal, A.M. Sengupta and S.R. Wadia,
            {\it Mod.\ Phys.\ Lett.\ }{\bf A6} (1991) 1465;
           J. Polchinski, {\it Nucl.\ Phys.\ }{\bf B362} (1991), 125.}
\REFSCON\dagr{U.H. Danielsson and D.J. Gross,
           {\it Nucl.\ Phys.\ }{\bf B366} (1991), 3.}
\REFSCON\satafact{N. Sakai and Y. Tanii,
           Tokyo Inst.\ of Tech.\ and Saitama preprint
           TIT/HEP--173, STUPP--91--120 (1991),
           submitted to {\it Phys.\ Lett.\ }{\bf B}.}
\REFSCON\polch{J. Polchinski,
           {\it Nucl. Phys.\ }{\bf B324} (1989), 123;
           {\it Nucl.\ Phys.\ }{\bf B346} (1990), 253;
           S.R. Das, S. Naik and S.R. Wadia,
           {\it Mod.\ Phys.\ Lett.\ }{\bf A4} (1989), 1033;
           S.R. Das and A. Jevicki,
           {\it Mod.\ Phys.\ Lett.\ }{\bf A5} (1990), 1639.}
\REFSCON\sata{N. Sakai and Y. Tanii,
           {\it Int.\ J. of Mod.\ Phys.\ }{\bf A6} (1991), 2743;
           I.M. Lichtzier and S.D. Odintsov,
           {\it Mod.\ Phys.\ Lett.\ }{\bf A6} (1991), 1953.}
\REFSCON\berkl{M. Bershadsky and I.R. Klebanov,
           {\it Phys.\ Rev.\ Lett.\ }{\bf 65} (1990), 3088;
           {\it Nucl.\ Phys.\ }{\bf B360} (1991), 559.}
\REFSCON\polyakov{A.M. Polyakov,
           {\it Mod.\ Phys.\ Lett.\ }{\bf A6} (1991), 635.}
\REFSCON\dejero{K. Demeterfi, A. Jevicki and J.P. Rodrigues,
           {\it Nucl.\ Phys.\ }{\bf B362} (1991), 125;
           {\it Nucl.\ Phys.\ }{\bf B365} (1991), 499.}
\REFSCON\witten{E. Witten, Princeton preprint
           IASSNS-HEP-91/51 (1991).}
\REFSCON\klpo{I.R. Klebanov and A.M. Polyakov,
           Princeton preprint PUPT-1281 (1991).}
\REFSCON\miya{D. Minic and Z. Yang,
           Texas preprint UTTG--23--91 (1991).}
\REFSCON\dfkulong{P. Di Francesco and D. Kutasov,
           Princeton preprint PUPT--1276 (1991).}
\REFSCON\kac{V.G. Kac, in {\it Group Theoretical Methods in Physics},
           eds.\ W. Beiglbock et al., Lecture Notes in Physics vol.\ 94
           (Springer-Verlag, 1979).}
\refsend
%
%%%%%%%%%%%%%%%%%%%%%%%%%%%%%%%%%%%%%%%%%%%%%%%%%%%%%%%%%%%%%%%%%%%%%%%
%
\Pubnum={TIT/HEP--179 \cr STUPP--91--122}  % November, 1991
\titlepage
\title{\bf Operator Product Expansion and Topological States in
$ c \! = \! 1$ Matter Coupled to 2-D Gravity}
\author{Norisuke Sakai}
\address{Department of Physics, Tokyo Institute of Technology \break
         Oh-okayama, Meguro, Tokyo 152, Japan}
\andauthor{Yoshiaki Tanii}
\address{Physics Department, Saitama University \break
         Urawa, Saitama 338, Japan}
\abstract
Factorization of the $N$-tachyon amplitudes in two-dimensional $c=1$
quantum gravity is studied by means of the operator product expansion of
vertex operators after the Liouville zero mode integration.
Short-distance singularities between two tachyons with opposite
chiralities account for all singularities in the $N$-tachyon amplitudes.
Although the factorization is valid, other possible short-distance
singularities corresponding to other combinations of vertex operators
are absent since the residue vanishes.
Apart from the tachyon states, there are infinitely many topological
states contributing to the intermediate states.
This is a more detailed account of our short communication on the
factorization.
\endpage
%
%%%%%%%%%%  Section 1  %%%%%%%%%%%%%%%%%%%%%%%%%%%%%%%%%%%%%%%%%%%%%%%
%
\chapter{Introduction}
Recent studies in matrix models \refmark{\brka), \gmil} have
made significant progress in  the nonperturbative treatment of
two-dimensional quantum gravity and string theory.
To make further progress, it is important to clarify these nonperturbative
results of matrix models from the viewpoint of
the usual continuum approach of the two-dimensional quantum gravity.
The most standard way of treating the continuum theory of quantum
gravity in two-dimensions is the so-called Liouville
theory\rlap.\refmark{\dika) - \kitarev}
Several works have recently appeared to compute the correlation
functions on the sphere topology using the Liouville
theory and the technique of analytic
continuation\rlap.\refmark{\gouli) - \taya}
These results are consistent
with those of matrix models\rlap.\refmark{\kostov) - \dagr}
So far only conformal field theories with central charge
$ c \le 1$  have been successfully coupled to quantum gravity.
\par
In a recent publication\rlap,\refmark{\satafact}
we have reported the result of the
factorization analysis of the $c=1$ quantum gravity.
The purpose of this paper is to give a more detailed account of
the analysis for understanding the factorization of
$c=1$  quantum gravity in terms of the short-distance singularities
arising from the operator product expansion (OPE) of vertex
operators.
\par
The $c = 1$ case is the richest and the most interesting.
It has been observed that this theory can be regarded effectively
as a critical string theory in two
dimensions, since the Liouville field zero
mode provides an additional ``time-like'' dimension besides the
obvious single spatial dimension given by the zero mode of the $c=1$
matter\rlap.\refmark{\polch}
We have a physical scalar particle corresponding to the center of mass
motion of the string.
Though it is massless, it is still referred
to as a ``tachyon'' following
the usual terminology borrowed from the critical string theory.
Since there are no transverse directions, the continuous (field)
degrees of freedom are exhausted by the tachyon field.
In fact, the partition function for the torus topology
was computed in the Liouville theory, and was found to give precisely
the same partition function as the tachyon field
alone\rlap.\refmark{\sata), \berkl}
However, there are indications of the existence of other
discrete degrees of freedom in the $ c=1$  quantum gravity.
Firstly, the correlation functions obtained in the matrix model
exhibit a characteristic singularity structure\rlap.\refmark{\grklne}
In the continuum approach of the Liouville theory, Polyakov has
observed that special states with discrete momenta and
Liouville energies can produce such poles, and has called these
operators co-dimension two operators\rlap.\refmark{\polyakov}
Moreover the two-loop partition function has been computed in a
matrix model and evidence has been noted for the occurrence of these
topological sates\rlap.\refmark{\dejero}
Most recently, the symmetry governing such topological states are
beginning to be understood\rlap.\refmark{\witten), \klpo}
It is clearly of vital importance to pin down the role played by
these topological states as much as possible.
In the critical string theory, the particle content of the theory and
unitarity have been most clearly revealed through the factorization
analysis of scattering amplitudes.
On the other hand, the factorization and unitarity of the Liouville
theory have not yet been well understood.
\par
Since we are interested in the short distance singularities, we consider
correlation functions on a sphere topology only.
We find that the singularities of the amplitudes can be
understood as short-distance singularities of two vertex operators
of tachyons with opposite chiralities.
We also find that the other possible short-distance singularities
corresponding to other combinations of tachyons are absent since the
residue vanishes.
It is found that infinitely many discrete states contribute
to the intermediate states of the factorized amplitudes,
apart from the tachyon states.
These are the co-dimension two operators of Polyakov
\refmark{\polyakov} and presumably are
topological in origin.
We have also explicitly constructed some of the topological states.
\par
Recently we have received two papers where other groups have discussed
the subject partly overlapping with
ours\rlap.\refmark{\miya), \dfkulong}
Their results are consistent with ours.
In particular, the paper by Di Francesco
and Kutasov has shown the decoupling of various topological states
in certain kinematical configurations
which was only partially demonstrated in our analysis.
%
%%%%%%%%%%  Section 2  %%%%%%%%%%%%%%%%%%%%%%%%%%%%%%%%%%%%%%%%%%%%%
%
\chapter{Correlation Functions on a Sphere}
In this paper we consider $c=1$  conformal matter realized by a
single bosonic field (string variable) $ X$  coupled to
two-dimensional quantum gravity
$$
\eqalign{
S_{\rm matter} [g, X]
    = & \, {1 \over 4\pi\alpha'} \int d^2 \xi \sqrt{g} g^{\alpha\beta}
        \partial_\alpha X \partial_\beta X, \cr
Z = \int {\cal D} g_{\alpha\beta} & {\cal D} X {1 \over V_{\rm gauge}}
      \, {\rm e}^{- S_{\rm matter} [g, X]},
}\eqn\pathintegral
$$
where $ \alpha'$  is the ``Regge slope parameter'' and
$ V_{\rm gauge}$  is the volume of the group of diffeomorphisms.
This is the so-called ``noncritical string'' in one dimension.
We use the method of Ref.\ \Prefmark{\dika}.
In this approach the metric is parametrized by a
diffeomorphism $f$ and the Liouville field $\phi$ representing the
freedom of local Weyl rescaling
$$
f^\ast g_{\alpha \beta}
= {\rm e}^\phi \hat g_{\alpha \beta}(\tau),
\eqn\metricdif
$$
where $\hat g_{\alpha \beta} (\tau) $ is a reference metric that
depends on the moduli parameters  $\tau$ of the Riemann surface.
The diffeomorphism invariance allows us to choose the conformal gauge.
On a sphere there is no moduli and the gauge condition is
$$
g_{\alpha \beta} = {\rm e}^\phi \hat g_{\alpha \beta}.
\eqn\metric
$$
In the following we shall set $ \alpha'=2$ for convenience.
The matter action in the conformal gauge \metric\ is
$$
S_{\rm matter} [\hat g, X]= {1 \over 2\pi} \int d^2 z
              \partial X \bar \partial X,
\eqn\xaction
$$
where we have used a holomorphic variable $z=\xi^1+i\xi^2$ and
an anti-holomorphic variable $\bar z=\xi^1-i\xi^2$.
If we want to consider $c<1$ case,
we need to introduce additional terms with a parameter
$\alpha_0 = - \sqrt{(1-c)/12}$
$$
S_{\rm matter, \alpha_0} [\hat g, X]=
{i \alpha_0 \over 4\pi} \int d^2 z \sqrt{\hat g} \hat R X
+ {i \alpha_0 \over 2\pi} \int ds \hat k X,
\eqn\adaction
$$
where $\hat R$ is the curvature of the two-dimensional surface with
the metric $\hat g_{\alpha \beta}$ and
$\hat k$ is the geodesic curvature along the boundary of the
surface parametrized by $s$.
After converting the Liouville field path integral measure into
the translationally invariant measure for a usual scalar
field\rlap,\refmark{\dika} we can treat the Liouville field almost
as a free field except for the nontrivial dynamics of the zero mode
due to the cosmological term with the cosmological constant
$\mu$\rlap.\refmark{\seirev), \poltalk}
The Liouville action is given by the following action involving
parameters $Q$, $\alpha$ and $\mu$
\refmark{\dika) - \kitarev}
$$
\eqalign{
S_{\rm L}[\hat g, \phi] = \, &
{1 \over 8\pi} \int d^2 z \sqrt{\hat g} \left( \hat g^{\alpha \beta}
\partial_\alpha \phi \partial_\beta \phi
- Q \hat R \phi + 8 \mu \, {\rm e}^{\alpha \phi} \right)
-{Q \over 4\pi} \int ds \hat k \phi \cr
= \, & {1 \over 2\pi} \int d^2 z  \left(
\partial \phi \bar \partial \phi
- {Q \over 4}\hat R \phi + 2 \mu \, {\rm e}^{\alpha \phi} \right)
-{Q \over 4\pi} \int ds \hat k \phi.
}
\eqn\liouville
$$
The correlation functions of $X$ and $\phi$ are given by
$$
\VEV{X(z,\bar z) X(w,\bar w)}=\VEV{\phi(z,\bar z) \phi(w,\bar w)}
=-{\rm log}|z-w|^2.
\eqn\xcorrfunc
$$
The energy-momentum tensor is given by
$$
T(z) = -{1 \over 2} (\partial X)^2 -{1 \over 2} (\partial \phi)^2
- {1 \over 2} Q \partial^2\phi.
\eqn\holemtensor
$$
If we want to consider $c<1$ case with $\alpha_0\not =0$ in
Eq.\ \adaction, we should add $i\alpha_0 \partial^2 X$ to $T(z)$.
\par
The parameter $Q$ is fixed by requiring \refmark{\dika) - \kitarev}
that the theory
does not depend on a choice of the reference metric
$\hat g_{\alpha \beta}$. One finds
in the present case of $c=1$
$$
 Q = 2 \sqrt{2}.
\eqn\qvalue
$$
If we want to consider the $c<1$ case with $\alpha_0\not=0$
in Eq.\ \adaction,
we should use $Q=\sqrt{(25-c)/3}=2\sqrt{2+\alpha_0^2}$ instead.
The other parameter $\alpha$ is determined by requiring
that the naive cosmological term operator
${\rm e}^{\alpha \phi}$ is a primary field of dimension $(1, 1)$
$$
-{1 \over 2} \alpha (\alpha+Q) = 1.
\eqn\alphacod
$$
In the present case of a one-dimensional string ($c=1$), one finds
$$
\alpha= - \sqrt{2}.
\eqn\paramet
$$
If we consider the $c<1$ case,
we should use $\alpha_{+}=-Q/2+|\alpha_0|$ instead.
\par
We have introduced a naive cosmological term
$8 \mu \, {\rm e}^{\alpha \phi} $.
Although the naive cosmological term operator ${\rm e}^{\alpha\phi}$
has conformal weight $(1, 1)$, it has been argued that the operator
does not correspond to a genuine local operator in the case of $c=1$
because of the Liouville zero mode path
integral\rlap.\refmark{\seirev} The correct cosmological
term operator in the $ c=1$  case is given by
$ \phi \, {\rm e}^{\alpha\phi}$  rather than
$ {\rm e}^{\alpha\phi}$.
We can obtain the correlation functions with the correct cosmological
term operator $\mu_r \phi \, {\rm e}^{\alpha \phi}$ from those with
the naive cosmological term operator $\mu \, {\rm e}^{\alpha\phi}$
by the following limiting procedure:
we replace $\alpha$ by $(1-\epsilon)\alpha$ and $\mu$ by
$\mu_r/(2 \epsilon)$ and take
the $\epsilon \rightarrow 0$ limit\rlap.\refmark{\dfku),\sataco}
This procedure is guaranteed to give the correct result,
since the naive cosmological terms decouple from the correlation
functions as we will explain later.
\par
The gravitationally dressed tachyon vertex operator $O_p$ with momentum
$ p$  is given by
$$
O_p = \int d^2z \sqrt{\hat g} \; {\rm e}^{i p X} \,
      {\rm e}^{\beta(p)\phi}.
\eqn\vertop
$$
We see that the Liouville zero mode can be regarded as
an ``imaginary time'' and the exponent $ \beta(p)$  as ``energy''.
The gravitational dressing of the tachyon vertex operator is
determined by requiring that the dressed operator has conformal
weight $(1, 1)$:
$$
{1 \over 2} p^2 - {1 \over 2} \beta (\beta+Q) = 1.
\eqn\onshellcod
$$
There are two solutions for the Liouville energy
$\beta(p)=-{Q \over 2} \pm |p|$.
It has been argued \refmark{\seirev), \poltalk} that the Liouville
zero mode path integral is well-defined only if $\beta>-Q/2$.
Hence we choose in the case of $c=1$
$$
\beta(p) = - \sqrt{2} + \; |p|.
\eqn\betatach
$$
For $c<1$, $p^2/2$ in Eq.\ \onshellcod\ should be replaced by
$p(p-2\alpha_0)/2$ and the solution becomes
$\beta(p)=-{Q \over 2} \pm |p-\alpha_0|$.
\par
The $N$-point correlation function of the tachyon vertex operators
\vertop\ on the sphere without boundary is given by the path integral
$$
\eqalign{
\VEV{O_{p_1} \cdots O_{p_N}}
= & \int {{\cal D} X {\cal D} \phi  \over V_{SL(2, {\bf C})}} \;\,
    {\rm e}^{- S_{\rm matter}[\hat g, X]
    - S_{\rm L}[\hat g, \phi]} \;\, O_{p_1} \cdots O_{p_N} \cr
= & \int \prod_{i=1}^N \left[d^2 z_i\sqrt{\hat g}\right] \,
    {1 \over V_{SL(2, {\bf C})}}
     \VEV{{\rm e}^{ip_1 X(z_1)} \cdots {\rm e}^{ip_N X(z_N)}}_X \cr
  &  \times \VEV{{\rm e}^{\beta_1 \phi(z_1)} \cdots
     {\rm e}^{\beta_N \phi(z_N)}}_\phi,
}\eqn\nptint
$$
where $V_{SL(2,\bf C)}$
is the volume of the $SL(2,{\bf C})$ group which is generated by
the conformal Killing vectors on the sphere
$$
V_{SL(2, {\bf C})} = \int {d^2z_a d^2z_b d^2z_c \over
                     |z_a-z_b|^2 |z_b-z_c|^2 |z_c-z_a|^2}.
\eqn\slc
$$
In Eq.\ \nptint, we have omitted powers of the string coupling constant
$g_{\rm st}^{-2}$ for the sphere topology, since we will deal with
correlation functions on the sphere only.
It is convenient to separate the zero modes in the path integral:
the zero mode $\phi_0$ of the Liouville
field  $(\phi = \phi_0 + \tilde \phi)$
and the zero mode $X_0$ of the matter field $(X = X_0 + \tilde X)$
and perform the zero mode integrations first.
The integration over the zero mode $\phi_0$ of the Liouville
field gives
$$
\eqalign{
\VEV{\prod_{i=1}^N {\rm e}^{\beta_i \phi}}_{\phi}
& = \int {\cal D} \phi \, {\rm e}^{- S_{\rm L} [\hat g, \phi]} \;
    {\rm e}^{\beta_1 \phi} \cdots {\rm e}^{\beta_N \phi} \cr
& = {1 \over -\alpha} \Gamma(-s)
    \int {\cal D} \tilde \phi \;
    {\rm e}^{- S_{{\rm L}, 0} [\hat g, \tilde \phi]}
    \left( {\mu \over \pi} \int d^2 w
    \sqrt{\hat g} \, {\rm e}^{\alpha \tilde \phi} \right)^s
    \prod_{i=1}^N {\rm e}^{\beta_i \tilde \phi} ,
}\eqn\zeroint
$$
where $s$ is given for the sphere as
$$
s=- {1 \over \alpha} \left( Q + \sum_{i=1}^N\beta_{i} \right).
\eqn\es
$$
The non-zero mode part of the Liouville action $S_{{\rm L}, 0}$
is given by the free field action.
Integration over the matter zero mode $X_0$ gives the momentum
conservation
$$
\sum_{j=1}^N p_j = 0.
\eqn\momcon
$$
It is convenient to introduce two-momenta for tachyons
${\bf p}_j = (p_j, - i \beta_j)$, for the naive cosmological terms
${\bf q} = (0,-i\alpha)$  and for the source
$ {\bf Q}=(0,-iQ)$ respectively.
The definition of $s$ together with the momentum conservation allow
us to write down ``energy-momentum conservation'' using two-momenta
$$
\sum_{j=1}^N {\bf p}_j + s {\bf q} + {\bf Q} =0.
\eqn\emocon
$$
Let us note that, in the case of $c<1$, the source two-momentum
becomes ${\bf Q}=(-2\alpha_0,-iQ)$ and the momentum conservation is
modified to $ \sum_{j=1}^N p_j = 2\alpha_0$. Hence the two-momentum
conservation \emocon\ is unchanged.
Therefore we obtain the $N$-point correlation function after the zero
mode integration
$$
\eqalign{
&\VEV{ \prod_{j=1}^N O_{p_j}} \cr
 = & \; 2\pi \delta \left( \sum_{j=1}^N p_j \right)
    {\Gamma(-s) \over -\alpha}
    \int {{\cal D} \tilde X {\cal D} \tilde \phi
    \over V_{SL(2, {\bf C})}} \;\,
    \prod_{j=1}^N O_{p_j}  \left( {\mu \over \pi} \int d^2 w \,
    \sqrt{\hat g} \, {\rm e}^{\alpha \tilde \phi(w)} \right)^s
    \; {\rm e}^{- S_0} \cr
 = & \; 2\pi \delta \left( \sum_{j=1}^N p_j \right)
    {\Gamma(-s) \over -\alpha}
    \int \prod_{i=1}^N \left[ \, d^2 z_i \, \sqrt{\hat g} \, \right]
    {1 \over V_{SL(2, {\bf C})}} \cr
  & \times \VEV{ \prod_{j=1}^N {\rm e}^{ip_j {\tilde X}(z_j)}}_{\tilde X}
    \VEV{ \left( {\mu \over \pi} \int d^2 w \,
    \sqrt{\hat g} \, {\rm e}^{\alpha \tilde \phi(w)} \right)^s
    \prod_{j=1}^N {\rm e}^{\beta_j \tilde \phi(z_j)}}_{\tilde\phi},
}\eqn\npnztint
$$
The expectation values with $\tilde\phi, \tilde X$  denote
the path integral over the non-zero modes $\tilde\phi, \tilde X$
with the free action
$$
S_0
= {1 \over 2\pi} \int d^2 z \left(
  \partial \tilde \phi  \bar \partial \tilde \phi
  + \partial \tilde X \bar \partial \tilde X \right).
\eqn\nzaction
$$
\par
For a non-negative integer $s$, we can evaluate the non-zero mode
$\tilde \phi$  integral by regarding the amplitude as a scattering
amplitude of $N$-tachyons and $s$  naive cosmological terms.
$$
\eqalign{
\VEV{\prod_{i=1}^N {\rm e}^{\beta_i \phi}}_{\phi}
= & \, {1 \over -\alpha} \Gamma(-s)
    \left( {\mu \over \pi} \right)^s
    \int \prod_{j=1}^s d^2 w_j
    \prod_{1 \le j < k \le s} |w_j-w_k|^{-2\alpha^2} \cr
  & \times \prod_{1 \le i < j \le N} |z_i-z_j|^{-2\beta_i \beta_j}
    \prod_{i=1}^N \prod_{j=1}^s |z_i-w_j|^{-2 \beta_i \alpha}.
}\eqn\liouvint
$$
After performing the path integral over $X$ and fixing the
$ SL(2, {\bf C})$  gauge invariance ($ z_1=0, z_2=1, z_3=\infty$),
we obtain an integral representation for
the $ N$-tachyon amplitude.
It is convenient to factor out the momentum conservation and to define
the normalized amplitude $\tilde A$
$$
\VEV{ \prod_{j=1}^N O_{p_j} }
= \; 2\pi \delta \left( \sum_{j=1}^N p_j \right)
    {1 \over -\alpha} \Gamma(-s) \tilde A (p_1, \cdots, p_N).
\eqn\tildea
$$
The normalized $N$-tachyon amplitude is given by
$$
\eqalign{
  \tilde A
= &  \left({\mu \over \pi}\right)^s
   \int \prod_{i=4}^N d^2 z_i \prod_{j=1}^s d^2 w_j
   \prod_{i=4}^N \left( |z_i|^{ 2 {\bf p}_1 \cdot {\bf p}_i}
    \, |1-z_i|^{2 {\bf p}_2 \cdot {\bf p}_i} \right)
   \prod_{4 \le i < j \le N}
    |z_i-z_j|^{ 2 {\bf p}_i \cdot {\bf p}_j} \cr
 &  \times
    \prod_{i=4}^N \prod_{j=1}^s
    |z_i-w_j|^{2 {\bf p}_i \cdot {\bf q}}
    \prod_{j=1}^s \left( |w_j|^{2 {\bf p}_1 \cdot {\bf q}}
    \, |1-w_j|^{2 {\bf p}_2 \cdot {\bf q}} \right)
    \prod_{1 \le j < k \le s} |w_j-w_k|^{2 {\bf q} \cdot {\bf q}}.
}\eqn\nptzint
$$
\par
In spite of the non-analytic relation \betatach\ between
energy $\beta$ and momentum $p$, we need to continue analytically
the formula into general complex values of momenta
in order to explore the singularity structure.
Therefore it is convenient to define the chirality of tachyons:
the tachyon has positive (negative) chirality if the tachyon
energy-momentum satisfies $(\beta+\sqrt2)/p =1(-1)$ irrespective of
the actual values of momentum\rlap.\refmark{\polyakov}
It seems to us that the operators with $\beta<-\sqrt2$
in Eq.\ \nptint\ are free from the trouble
noted in  Refs.\ \Prefmark{\seirev), \poltalk} since the Liouville
zero mode $\phi_0$  has already been integrated out.
The physical values of momenta are reached by analytic continuation
in $s$, since $s$  is related to other momenta through \es.
For generic physical values of momenta $(s\not =0)$, one finds a finite
result for the $N$-tachyon amplitudes (both
$\VEV{\prod_{j=1}^N O_{p_j}}$ and $\tilde A (p_1, \cdots, p_N)$
are finite).
However, the result is different in different chirality
configurations, since the amplitude is non-analytic in momenta.
\par
Let us consider the kinematical configuration where all tachyons
 except one have the same chirality.
If $ p_1$  has negative chirality and the rest $ p_2,\cdots,p_N$
positive chirality, momentum conservation reads
$$
p_1+p_2+ \cdots +p_N=0
\eqn\nmomcon
$$
and energy conservation dictates that ($\alpha =-\sqrt2$)
$$
-p_1+p_2+ \cdots +p_N= \sqrt{2}(N+s-2).
\eqn\nenecon
$$
Thus one obtains the kinematical constraints
$$
p_1=-{N+s-2 \over \sqrt{2}}, \qquad
\beta_1={N+s-4 \over \sqrt{2}}.
\eqn\nkin
$$
It has been shown that the $N$-tachyon amplitude is given in this
kinematical configuration by \refmark{\gouli) - \dotsenko}
$$
\tilde A (p_1, \cdots, p_N)
=  {\pi^{N-3} [\mu\Delta (-\rho)]^s \over \Gamma(N+s-2)}
   \prod_{j=2}^N \Delta(1 - \sqrt{2} p_j),
\eqn\nptfin
$$
where  $\Delta (x)=\Gamma(x)/ \Gamma(1-x)$.
The regularization parameter $\rho$ is given by $\rho = - \alpha^2 /2$
and is eventually set equal to $ -1$  after analytic continuation
(in the central charge $c$).
We see immediately that the insertion of the naive cosmological term
operator always gives vanishing correlation functions, since
$\Delta (-\rho)=\Gamma(-\rho)/ \Gamma(1+\rho)$ vanishes at $\rho=-1$.
This decoupling of the naive cosmological term operator guarantees the
validity of our procedure in computing the correlation function with the
correct cosmological term operators $\phi \, {\rm e}^{\alpha\phi}$:
we replace $\alpha$ by $(1-\epsilon)\alpha$ and $\mu$ by
$\mu_r/(2 \epsilon)$ and take
the $\epsilon \rightarrow 0$
limit\rlap.\refmark{\dfku), \sataco} In effect, we should just
replace the combination $\mu\Delta(-\rho)$ by the correct
(renormalized) cosmological constant $\mu_r$.
For the case of negative chirality for $p_1$ and positive
chirality for the remaining $p_2, \cdots, p_N$, the momentum of
$p_1$ is fixed because of the kinematical constraints \nkin.
As a function of the other momenta $p_2,\cdots,p_N$,
the $N$-tachyon amplitude exhibits singularities at
$$
p_j = {n+1 \over \sqrt2},
\quad j = 2, \cdots N; \;\; n = 0, 1, 2, \cdots,
\eqn\poles
$$
but has no singularities in other combinations of momenta contrary
to the dual amplitudes in the critical string theory.
The first pole will be shown to correspond to tachyon as an intermediate
state in the next section.
Other higher level poles for $ n = 1, 2, \cdots$ will be shown to
correspond to topological states as argued by several
people\rlap.\refmark{\grklne), \polyakov}
\par
Let us examine other kinematical configurations.
The amplitudes with one tachyon of positive chirality and the rest
negative are given by changing the sign of $p_j$ in Eq.\ \nptfin.
On the other hand, if each chirality has two or more tachyons, the
normalized amplitude $\tilde A$  is finite for generic momenta
but has the factor $1/\Gamma(-s)$.
Hence $\tilde A$ vanishes for more than two tachyons in each chirality,
when we consider non-negative integer $s$  in the following.
This property has been explicitly demonstrated for the four- and
five-tachyon amplitudes
in the Liouville theory\rlap,\refmark{\dfku), \polyakov}
and has been argued to be a general
property using the matrix model\rlap.\refmark{\grklne}
Therefore we take it for granted that the tachyon scattering amplitudes
$ \tilde A$  vanish for non-negative integer $s$, unless
there is only one tachyon in either one of the chiralities.
\par
Let us note that our assertion is consistent with the argument for
vanishing S-matrix by Gross and Klebanov\refmark{\grklne}:
they absorb the $ \Delta(1 \pm \sqrt{2} \, p)$  factor for the
individual momenta into a renormalization factor of tachyon vertex
operators.
This momentum dependent renormalization factor
is harmless if the momenta are at some generic values
$(p\not=(n+1)/\sqrt2)$.
However, the two-dimensional kinematics forces one of the
renormalization
factors to be infinite, if there is only one tachyon in either one
of the chiralities\rlap,\refmark{\dfku) - \sataco}
since the kinematical constraints
fix the momentum of the tachyon to be at one of the poles \poles .
In the case of
negative chirality for $ p_1$ and positive chirality for the rest
$p_2,\cdots,p_N$, the momentum $p_1$ is fixed to be \nkin .
Because of this infinite renormalization,
the renormalized amplitudes of Gross and Klebanov vanish
even if there is only one tachyon in either one of the chiralities.
\par
Let us suppose that we are interested in the case of vanishingly small
values of the cosmological constant $\mu$.
We note that, even if the cosmological constant $\mu$ is infinitesimal,
the cosmological term $\mu{\rm e}^{\alpha \phi}$ can become
arbitrarily large for sufficiently large negative values of the
Liouville field $\phi$ ($\alpha<0$ in our convention).
Therefore the Liouville field in the path integral is suppressed for
large negative values.
Since the large positive values of the Liouville field should be cut
off as an ultraviolet or short-distance cut-off,
the tachyon field space is restricted to a large
but finite volume (proportional to $\ln \mu$)
\rlap.\refmark{\seirev), \poltalk}
It is important to remember that we cannot neglect the cosmological
constant completely even if it is infinitesimally small.
On the other hand, the contribution to the amplitudes proportional to
$\ln \mu$ is given by the tachyon interaction in the bulk and hence
is insensitive to the details of the cut-off of the Liouville field
space. The nonlinearity of the Liouville interactions remains only
in the form of the restricted field space.
We are precisely interested in this bulk interaction of tachyons.
The finite values of the normalized scattering amplitudes
$\tilde A$ correspond to a divergent correlation function at
$s=$ positive integers.
Since the correlation functions are given by $\tilde A$ multiplied by
$(\mu)^s\Gamma(-s)$, the divergent correlation functions are more
properly interpreted as the logarithmically divergent contribution as
we let $\mu\rightarrow 0$.
Therefore the finite tachyon scattering amplitudes $\tilde A$ at
$s=$ non-negative integers represent the tachyon interaction
proportional to the volume of the Liouville field
space\rlap.\refmark{\grklne), \polyakov), \dfku}
Hence they are often called the bulk or resonant amplitudes.
\par
%
%%%%%%%%%%  Section 3  %%%%%%%%%%%%%%%%%%%%%%%%%%%%%%%%%%%%%%%%%%%%%
%
\chapter{Higher Level Operators}
Since the short-distance singularities should come from terms in
the OPE, we first examine the operators responsible for these
singularities. We will consider the OPE of vertex operators
after the zero mode integration of $\phi$ and $X$ in Eq.\ \npnztint.
Therefore the operators we consider in this section
consist of only non-zero mode ${\bf X} = (\tilde\phi, \tilde X)$,
which have the free action \nzaction.
The simplest operator is $ {\rm e}^{i {\bf p} \cdot {\rm X}}$,
which is the tachyon operator \vertop\ with zero modes omitted.
For higher levels, it has been pointed out that there are only
null states at generic values of
momenta\rlap.\refmark{\polyakov), \grklne}
However, there are exceptional values of momenta where the null states
degenerate and new primary states emerge as a result.
These new primary states are called co-dimension two
states by Polyakov\rlap,\refmark{\polyakov} and special states,
topological states or discrete states by other
people\rlap.\refmark{\grklne), \berkl}
We can construct vertex operators for these topological states
in the following way.
\par
The energy-momentum tensor for $ {\bf X}$  is given by
$$
\eqalign{
T(z) & = - {1 \over 2} \partial {\bf X} \cdot \partial {\bf X}
      - {1 \over 2} i {\bf Q} \cdot \partial^2 {\bf X} \cr
  & = - {1 \over 2} \partial \tilde X \partial \tilde X
      - {1 \over 2} \partial \tilde \phi \, \partial \tilde \phi
      - {1 \over 2} Q \partial^2 \tilde \phi
}\eqn\emtensor
$$
and the anti-holomorphic component given by \emtensor\
with $\partial$ replaced by $\bar\partial$.
They satisfy the Virasoro algebra of the central charge 26.
We should construct the field of conformal weight $ (1, 1)$
with respect to this energy-momentum tensor by taking linear
combinations of monomials of derivatives of $ {\bf X}$
multiplied by $ {\rm e}^{i{\bf p} \cdot {\bf X}}$.
The operator at level $ n$  has $ n$  $ \partial$'s and
$ n$  $ \bar\partial$'s for each monomial.
The condition for the conformal weight to be $ (1,1)$  at level $ n$  is
$$
{1 \over 2} {\bf p} \cdot ({\bf p} + {\bf Q}) + n = 1.
\eqn\polen
$$
We should note that there are two branches of the solution for the
condition \polen\
$$
\beta=-\sqrt2\pm\sqrt{p^2+2n}.
\eqn\enen
$$
Seiberg has noted that the vertex operator with $\beta>-\sqrt2$
gives an ill-defined integration over the Liouville zero
mode\rlap.\refmark{\seirev), \poltalk}
Hence the lower sign in Eq.\ \enen\ is forbidden by this condition.
However, we consider both cases here, since we are considering
vertex operators consisting of non-zero mode $\tilde\phi$  only.
In fact, the $s$  naive cosmological terms are a result of
the Liouville zero mode $\phi_0$ integration.
We shall call the upper sign solution
S- (Seiberg) type and the lower sign A- (anti-Seiberg) type.
\par
At level $n=1$  the general form of the vertex operator with
one $ \partial$  and one $ \bar\partial$  is
$$
V = \zeta_{\mu\nu} \partial {\bf X}^\mu \bar\partial {\bf X}^\nu
    {\rm e}^{i{\bf p} \cdot {\bf X}}.
\eqn\levelone
$$
For this operator to be a primary field of the unit conformal weight,
the OPE with the energy-momentum tensor \emtensor\ must be
$$
T(z) \, V(w) \sim {1 \over (z-w)^2} \, V(w)
                  + {1 \over z-w} \, \partial V(w),
\eqn\primary
$$
which gives the conditions on the polarization tensor
$$
({\bf p} + {\bf Q})^\mu \zeta_{\mu\nu} = 0
= \zeta_{\mu\nu} ({\bf p}+{\bf Q})^\nu
\eqn\polcondone
$$
and the on-shell condition \polen\ with $ n=1$.
A similar condition should also be satisfied for the anti-holomorphic
part in order for the operator to be a $(1,1)$ primary.
Solving these conditions we find only one primary field with
weight $ (1, 1)$  at generic values of momentum, i.e.\ $ p\not=0$
$$
V^{(1)}
= \, {\bf p} \cdot \partial {\bf X} \;
  {\bf p} \cdot \bar \partial {\bf X} \,
  {\rm e}^{i{\bf p} \cdot {\bf X}}
= - L_{-1} \bar L_{-1} {\rm e}^{i{\bf p} \cdot {\bf X}}.
\eqn\onenull
$$
The above state is clearly null.
However, the situation changes at $p=0$.
For the S-type, the operator vanishes at $p=0$.
Therefore we can construct a new operator by a limit
$$
V_{(1,1)}= \lim_{p\rightarrow 0}{V^{(1)} \over p^2}
= \partial X \bar \partial X.
\eqn\gravopt
$$
We easily find that this field is primary and not null.
This kind of a peculiar operator exists only at a discrete momentum.
Since there are two kinematical constraints to specify the state,
one for the energy and the other for the momentum, the state is called
co-dimension two\rlap.\refmark{\polyakov}
As for the A-type at $p = 0$, we find that the $(1, 1)$  operator
condition does not constrain the polarization tensor multiplying
the operator $ \partial {\bf X} \, \bar\partial {\bf X} \,
{\rm e}^{i{\bf p} \cdot {\bf X}}$.
Hence we again obtain a new primary field
$$
V'_{(1,1)}= \partial X \bar \partial X{\rm e}^{-2\sqrt2 \phi}.
\eqn\agravopt
$$
\par
At level two, the general form of operators is given by
(we present only the holomorphic part)
$$
V = \left( \zeta_{\mu} \partial^2 {\bf X}^\mu
    + \zeta_{\mu\nu} \partial {\bf X}^\mu \partial {\bf X}^\nu \right)
    {\rm e}^{i{\bf p} \cdot {\bf X}}.
\eqn\leveltwo
$$
The OPE \primary\ gives the conditions
$$
\eqalign{
\zeta_{\mu\mu} + i(3{\bf Q}+2{\bf p})^\mu \zeta_\mu & = 0, \cr
\zeta_\mu - i({\bf p}+{\bf Q})^\nu \zeta_{\nu\mu} & = 0
}
\eqn\polcondtwo
$$
and Eq.\ \polen\ with $ n=2$.
We find two independent solutions to these conditions
for the holomorphic part
$$
\eqalign{
V^{(2)} & = \left( L_{-2} + {3 \over 2} L_{-1}^2 \right)
            {\rm e}^{i {\bf p} \cdot {\bf X}}, \cr
V^{(3)} & = L_{-1} \left( { 1\over 4} i \left[
            ( 8 - {\bf p} \cdot {\bf Q} ) {\bf p} - 2 {\bf Q}
            \right] \cdot \partial {\bf X} \,
            {\rm e}^{i {\bf p} \cdot {\bf X}} \right).
}\eqn\twonull
$$
Both fields are null. At a special value of the momentum
these two operators are linearly dependent and we obtain
a topological state.
For instance, the (2,1) topological state of S-type is given by
$$
\eqalign{
V_{(2, 1)}
= & \lim_{p \rightarrow {1 \over \sqrt{2}}}
    {6 \sqrt{2} \over p - {1 \over \sqrt{2}}}
    ( V^{(2)} - V^{(3)} ) \cr
= & \, ( 13 \, \partial X \partial X - \partial \phi \partial \phi
    - 6 \, i \, \partial X \partial \phi - \sqrt{2} \, i \, \partial^2 X
    - \sqrt{2} \, \partial^2 \phi ) \,
    {\rm e}^{{1 \over \sqrt{2}} \, i \, (X - i \phi)}.
}\eqn\prmop
$$
We find exactly the same situation for the anti-holomorphic part.
We can continue to explore $(1,1)$ operators at higher levels similarly.
We expect that these $ (1,1)$  operators are null fields for generic
values of momenta, and that, at special values of momenta,
these null states are not linearly
independent, namely they degenerate.
Then we obtain a new primary state from a limit of an appropriate
linear combination of these null states.
We expect to have both S-type and A-type topological states.
\par
There are other procedures to obtain topological states.
These states were found to originate from the gravitational dressing
of the primary states in the $ c=1$  conformal field theory which
create null descendants at
level $n$\rlap.\refmark{\dagr), \berkl), \kac}
The momentum $ p$  of the initial primary state and the level $ n$  are
specified by two positive integers $ (r, t)$
$$
p={r-t \over \sqrt2},\qquad
n=rt .
\eqn\rtstate
$$
Since the level $n$ corresponds to the $n$-th derivatives, we find that
the energy $\beta$  of the topological state is determined by the
$(1, 1)$ conformal weight condition \enen\
and is given by
$$
\beta={-2 \pm (r+t) \over \sqrt2}.
\eqn\rtenergy
$$
The upper (lower) sign corresponds to the S- (A-) type solution.
For example, $(r, t)=(1, 1)$  and $(2, 1)$  operators of S-type are
$$
\eqalign{
V_{(1, 1)} & = \partial X \bar\partial X, \cr
V_{(2, 1)} & = \left( \partial X \partial X
               + {1 \over \sqrt{2}} i \partial^2 X \right)
               \left( \bar\partial X \bar\partial X
               + {1 \over \sqrt{2}} i \bar\partial^2 X \right)
               {\rm e}^{{1 \over \sqrt{2}} (i X + \phi)}.
}\eqn\topologicalop
$$
We find that these operators coincide with our operators
\gravopt\ and \prmop\ respectively
up to only a certain amount of null operators.
\par
%
%%%%%%%%%%  Section 4  %%%%%%%%%%%%%%%%%%%%%%%%%%%%%%%%%%%%%%%%%%%%
%
\chapter{OPE and Factorization}
To understand the poles of the amplitudes in terms of
short-distance singularities in the OPE,
we shall consider the case of $ s=$ non-negative integers
by choosing the momentum configuration appropriately.
As we explained before, these amplitudes at non-negative integer
$s$  represent the so-called bulk or resonant
interactions\rlap.\refmark{\grklne), \polyakov}
The only nonvanishing $N$-tachyon amplitudes $\tilde A$ at
$s=$ non-negative integers
are for the kinematical configuration where one of the chiralities
have only a single tachyon and the rest opposite chirality.
Here we shall take the case of $s = 0$  for the $N$-tachyon
amplitude with only one negative chirality tachyon ($ p_1$), and
examine the $ s=$ positive integers case at the end.
\par
First we shall illustrate the origin of short-distance singularities
in the simplest context of the four tachyon
scattering amplitude with $ s=0$.
As in Eq.\ \nptzint\ we fix $ z_2=0,z_3=1,z_4=\infty$ and
set $ z_1=z$ to find
$$
\tilde A (p_1,\cdots,p_4)
 =  \int d^2 z \, |z|^{ 2 {\bf p}_1 \cdot {\bf p}_2}
     \, |1-z|^{ 2 {\bf p}_1 \cdot {\bf p}_3}.
\eqn\fourptzintfull
$$
The short-distance singularities corresponding to $ z_1\sim z_2$
($z\rightarrow 0$) can be exhibited by
expanding the integrand around $z=0$
$$
\eqalign{
\tilde A (p_1,\cdots,p_4)
& \approx \int_{|z| \le \epsilon} \!\! d^2 z \,
          |z|^{ 2 {\bf p}_1 \cdot {\bf p}_2}
          \left| \, \sum_{n=0}^{\infty}
          \biggl({\Gamma(1+{\bf p}_1 \cdot {\bf p}_3) \over n! \,
          \Gamma({\bf p}_1 \cdot {\bf p}_3-n+1)}\biggr)
          (-z)^n \, \right|^2 \cr
& \approx \sum_{n=0}^{\infty}
{\pi \over n+1+{\bf p}_1 \cdot {\bf p}_2}
          \biggl({\Gamma(1+{\bf p}_1 \cdot {\bf p}_3) \over n! \,
          \Gamma({\bf p}_1 \cdot {\bf p}_3-n+1)}\biggr)^2.
}\eqn\fourptzint
$$
We see that the so-called noncritical string of the $c=1$ quantum
gravity exhibits exactly the same type of short-distance singularities
as the familiar critical string theory.
The kinematics at the pole reflects the peculiarities of
two-dimensional physics
$$
\eqalign{
{\bf p}_1 & =(-\sqrt2,0), \qquad
{\bf p}_2=({n+1 \over \sqrt2},-i{n-1 \over \sqrt2}), \cr
{\bf p}_j & =(p_j,-i(-\sqrt2+p_j)) \qquad j=3,\, 4.
}\eqn\kinpole
$$
We can express the short-distance singularities in terms of these
momenta to find
$$
\tilde A (p_1,\cdots,p_4)
 \approx \sum_{n=0}^{\infty} {(-1)^n \over (n!)^2}
{\pi \over n+1-\sqrt2 \, p_2} \prod_{j=3}^4\Delta(1-\sqrt2 \, p_j).
\eqn\fourptsing
$$
This shows that all the singularities in $ p_2$  in the full amplitude
are correctly accounted for by these short-distance singularities
near $ z_1\sim z_2$.
Similarly, the singularities in $ p_3$  and $ p_4$  are accounted for
by the short-distance singularities for $ z_1 \sim z_3$ and
$z_1 \sim z_4$ respectively.
Hence we see that all the singularities in the amplitude are nothing
but the short-distance singularities associated with the opposite
chirality tachyons approaching each other.
\par
We can understand these short-distance singularities by means of
the OPE of two vertex operators
$$
:{\rm e}^{i{\bf p}_1 \cdot {\bf X}(z_1)}:
:{\rm e}^{i{\bf p}_2 \cdot {\bf X}(z_2)}: \,
\sim \sum_{n=0}^{\infty}\biggl({1 \over n!}\biggr)^2
|z_1-z_2|^{2{\bf p}_1 \cdot {\bf p}_2+2n} \, V_n (z_2).
\eqn\tacope
$$
The operators on the right hand side are given by
$$
\eqalign{
V_n & = \; :{\rm e}^{i{\bf p}_2 \cdot {\bf X}}
        \partial^n\bar\partial^n {\rm e}^{i{\bf p}_1
        \cdot {\bf X}}: \cr
    & = \; : ( - {\bf p}_1 \cdot \partial^n {\bf X} \;
        {\bf p}_1 \cdot \bar\partial^n {\bf X} + \cdots ) \,
        {\rm e}^{i {\bf p} \cdot {\bf X}} :,
}\eqn\opeop
$$
where ${\bf p} = {\bf p}_1 + {\bf p}_2$.
It can be shown that $ V_n$  is a primary field with a conformal
weight $ (1, 1)$  when $ {\bf p}$  satisfies  Eq.\ \polen.
Integration of Eq.\ \tacope\ by $ z_1$  over the region
$|z_1-z_2| \le \epsilon$  gives singularities in the momentum $ {\bf p}$
$$
\sum_{n=0}^\infty \left( {1 \over n !} \right)^2
{\pi \over {1 \over 2} {\bf p} \cdot ( {\bf p} + {\bf Q} ) + n - 1}
V_n (z_2).
\eqn\intope
$$
To discuss the OPE in a general context, we now consider
the $N$-tachyon amplitude where $p_1$ has negative chirality and
$p_2,\cdots, p_N$ have positive chirality.
{}From the energy-momentum conservation \nmomcon\ and \nenecon\
for this kinematical configuration,
the two-momentum of the tachyon 1 takes a fixed value
$$
{\bf p}_1 = \left( -{N-2 \over \sqrt{2}},
-i{N-4 \over \sqrt{2}} \right).
\eqn\negtwomom
$$
This kinematical constraint implies for the intermediate state
momentum ${\bf p}={\bf p}_1+{\bf p}_2$
$$
 {1 \over 2} {\bf p} \cdot ( {\bf p} + {\bf Q} ) + n - 1 =
(N-3)(1-\sqrt{2} p_2) + n.
\eqn\intprod
$$
At the pole $p_2=1/\sqrt2$  ($n=0$),
the intermediate state momentum becomes
$$
{\bf p} = \left( -{N-3 \over \sqrt{2}},
-i\left(-\sqrt2+{N-3 \over \sqrt{2}}\right) \right).
\eqn\inttwomom
$$
Hence we find that the pole at $p_2=1/\sqrt2$ ($n=0$)
in Eq.\ \intope\ is due to the tachyon intermediate state $ V_0$ of
negative chirality.
The higher level poles ($ n \ge 1$) are due to the topological states
in $ V_n \; (n \ge 1)$  that we have discussed in the previous section.
These topological state operators can take various appearances depending
on the amount of null states.
If we let
$$
 N-3 = t, \qquad n = r t,
\eqn\rtnnkin
$$
then the singular factor in Eq.\ \intope\ for $ n \ge 1$  becomes
$$
{1 \over {1 \over 2} {\bf p} \cdot ( {\bf p} + {\bf Q} ) + n - 1}
= {1 \over t (r+1-\sqrt{2} p_2)}.
\eqn\topsing
$$
Hence we obtain the pole
$$
 p_2 \rightarrow {r+1 \over \sqrt{2}} .
\eqn\rtnnpkin
$$
In this case the intermediate state momentum becomes
$$
\eqalign{
{\bf p} & = \left( {r-t \over \sqrt{2}},
-i\left(-\sqrt2+{r+t \over \sqrt{2}}\right) \right) \cr
& = \left( {r-t \over \sqrt{2}},
-i\left(-\sqrt2+\sqrt{\left({r-t \over \sqrt{2}}\right)^2+rt}
\right) \right).
}\eqn\inttwomom
$$
This two-momentum is precisely the two-momentum \rtstate\ and
\rtenergy\ of the topological state at level $n=rt$ of the S-type.
Therefore we find that the $(r, t)$ topological state
is contained in $ V_n$ if
the number of tachyons $N$ and the level of the intermediate state $n$
are specified by Eq.\ \rtnnkin\ and
that the operator must be of S-type.
Thus we find that each $(r, t)$ primary topological state
appears as an intermediate state of the $N$-tachyon amplitude at level
$n$ exactly once.
This together with the above tachyon singularity explains all
singularities in Eq.\ \poles.
\par
Let us discuss the residues of these short-distance singularities.
These poles can be associated with the intermediate
tachyon or topological states appearing in the OPE.
As illustrated in Fig.\ 1 the residues of these poles are given by a
product of two parts.
One of them, the right side blob in Fig.\ 1, has $ N-2$  tachyons
with incoming momenta ${\bf p}_3, \cdots, {\bf p}_N$, and an
intermediate particle with incoming momentum
${\bf p}={\bf p}_1+{\bf p}_2$, and is given by a kind of dual amplitude.
The other part has two tachyons with incoming momenta
${\bf p}_1, {\bf p}_2$  and the intermediate
particle as shown in the left side blob in Fig.\ 1.
First we examine the pole at $ p_2=1/\sqrt2$, namely at the lowest
level $(n=0)$.
This pole is due to the tachyon
intermediate state with negative chirality.
In fact we find that the residue of the pole $ p_2=1/\sqrt2$  is
precisely given by a product of the tachyon three point function and
the $ N-1$  tachyon amplitude with a single
(intermediate state) tachyon $ {\bf p}$  having negative chirality
and the rest $ {\bf p}_3, \cdots, {\bf p}_N$
having positive chirality
$$
\eqalign{
\tilde A(p_1, p_2, p_3, \cdots, p_N)
& \approx {\pi \over (N-3)(1-\sqrt2 p_2)}
  \tilde A(p, p_3, \cdots, p_N) \cr
& = \tilde A(p_1, p_2, -p)
  {2\pi \over {\bf p} \cdot ( {\bf p} + {\bf Q} ) -2}
  \tilde A(p, p_3, \cdots, p_N).
}\eqn\tacpol
$$
This shows that the factorization is valid similarly to critical
string theory. By symmetry, we can explain the lowest poles in
each individual momentum $ p_j=1/\sqrt2$  as the tachyon
intermediate state in the OPE of $ p_1$  and $ p_j$.
\par
For higher level poles, we explicitly evaluate the residue of
the short-distance singularities
up to $ p = 3 / \sqrt{2}$  and up to $ N = 5$.
For instance, the five tachyon amplitude has short-distance
singularities at $ p_2=2/\sqrt2$  and $ 3/\sqrt2$
$$
\tilde A (p_1, \cdots, p_5) \approx
\left[ - {\pi^2 \over 2(2 - \sqrt{2} \, p_2)}
+ {\pi^2 \over 8(3 - \sqrt{2} \, p_2)} \right]
\prod_{j=3}^5 \Delta(1-\sqrt2 \, p_j).
\eqn\singfive
$$
The residues of these poles correctly reproduce the residues
of the poles in the full amplitude.
Each term of Eq.\ \singfive\ can also be written as
$$
\tilde A(p_1, p_2; -p)
{2\pi \over {\bf p} \cdot ( {\bf p} + {\bf Q} ) + 2(n - 1)}
\tilde A(p; p_3, \cdots, p_N),
\eqn\factorization
$$
where $ n=1$  and $ 2$  for the first and the second terms respectively
and $ -p$  and $ p$  are the momenta of the intermediate particles.
We have explicitly verified that these amplitudes appearing in the
residue agree with those obtained from the three and $N-1$ point
amplitudes with one topological state of momentum $-p$ and $p$.
It is rather difficult to compute the short-distance singularities
explicitly to an arbitrary level except for the four-point amplitude
that we have already worked out in Eq.\ \fourptzint.
Therefore we content ourselves with the computation of lower level
singularities in explicitly demonstrating that the singularities of
the amplitudes all come from the short-distance singularities
of $ p_1$  and $ p_j$.
\par
The OPE suggests that there may be other short-distance singularities
in other combinations of momenta if one considers corresponding
combinations of vertex operators approaching to the same point.
For instance, short-distance singularities corresponding to
$ k$  vertex operators approaching each other, say $ z_1, \cdots, z_k$,
should give poles in $ p_2+\cdots+p_k$.
It is most convenient to fix reduced variables
$ u_j = (z_j-z_2) / (z_1-z_2)$  $ (j = 1, \cdots, k)$
in taking the short-distance limit
$ z_1 \rightarrow z_2$.
The amplitude exhibits short-distance singularities whose
residues are given by a product of two dual amplitudes (Fig.\ 2)
$$
\eqalign{
\tilde A(p_1, \cdots, p_N)
\approx & \, {1 \over V_{SL(2, {\bf C})}} \int_{|z_1-z_2|\le \epsilon}
  \!\!\!\!\! d^2 z_1 d^2 z_2
  \prod_{i=3}^k d^2 u_i \prod_{j=k+1}^N d^2 z_j \,
  |z_1 - z_2|^{{\bf p} \cdot {\bf p} + {\bf Q} \cdot {\bf p} - 4} \cr
& \times \prod_{1 \le i < j \le k}
  |u_i - u_j|^{2 \, {\bf p}_i \cdot {\bf p}_j}
  \prod_{i=1}^k \prod_{j=k+1}^N \left|
  1 + {z_1 - z_2 \over z_2 - z_j} \, u_i
  \right|^{2 \, {\bf p}_i \cdot {\bf p}_j}  \cr
& \times \prod_{i=k+1}^N |z_2 - z_i|^{2 \, {\bf p} \cdot {\bf p}_i}
  \prod_{k+1 \le i < j \le N}
  |z_i - z_j|^{2 \, {\bf p}_i \cdot {\bf p}_j}.
}\eqn\twodual
$$
The dual amplitude with the original variables
$ z_i$  ($ i = 2, k+1, \cdots, N$) has
$ N-k$  positive chirality tachyons $ {\bf p}_{k+1},\cdots,{\bf p}_N$
and the intermediate state particle $ {\bf p}$  (right side blob in
Fig.\ 2), whereas
the dual amplitude with the reduced variables
$ u_j$  ($ j = 3, \cdots, k$) has the
intermediate state particle $ -{\bf p}-{\bf Q}$  and $ k$  tachyons
${\bf p}_1,\cdots,{\bf p}_k$  whose chiralities are
positive except $ {\bf p}_1$
(left side blob in Fig.\ 2).
\par
It is important to clarify the kinematical constraints on the
intermediate states when the amplitudes are factorized.
In the left side blob in Fig.\ 2, the incoming momenta of tachyons
${\bf p}_1,\cdots,{\bf p}_k$ are balanced by the incoming momentum
$-{\bf p}=-{\bf p}_1-\cdots-{\bf p}_k$.
However, if we want to interpret
the left side blob as a dual amplitude
coming from a path integral with our action \xaction\ and \liouville,
we need to assign the incoming two-momentum of the intermediate particle
to be $- {\bf p} - {\bf Q}$, since the action dictates that
an external source two-momentum ${\bf Q}$ should be present.
Therefore if ${\bf p}$  is the intermediate state momentum flowing
into the right side blob in  Fig.\ 2, the corresponding momentum for
the dual amplitude of the left side blob should be regarded as
$-{\bf p}-{\bf Q}$.
Consequently, if the intermediate state is a tachyon, the chirality
of the tachyon for the left side blob
is the same as that of the tachyon for the right side blob
$$
{\bf p}=(p, -i(-\sqrt2\pm p)), \qquad
-{\bf p}-{\bf Q}=(-p, -i(-\sqrt2\pm(-p))).
\eqn\inttach
$$
Similarly, if the intermediate state is a topological state,
the type (S or A) of the intermediate state for the left side blob
is the opposite to that of the intermediate state for the right
side blob
$$
{\bf p}=(p, -i(-\sqrt2 \pm \sqrt{p^2+2n})), \quad
-{\bf p}-{\bf Q}=(-p, -i(-\sqrt2 \mp \sqrt{p^2+2n})).
\eqn\inttopo
$$
\par
In the present case, we have only a single tachyon with the negative
chirality whose two-momentum is determined by the two-dimensional
kinematics \negtwomom.
Since we pinch together the single negative chirality tachyon with the
positive chirality tachyons ${\bf p}_1,\cdots,{\bf p}_k$,
we find the intermediate particle to have the two-momentum
$$
{\bf p} = \left( {-N+2 \over \sqrt{2}}+\sum_{j=2}^k p_j,
-i\left({N-2-2k \over \sqrt{2}}+\sum_{j=2}^k p_j\right) \right).
\eqn\kinttwomom
$$
Since $k\le N-2$, it is tachyon if
$$
\sum_{j=2}^k p_j={k-1 \over \sqrt2}
\eqn\negtach
$$
and its chirality is always negative.
In this case, the left side blob has more than two tachyons
for each chirality.
On the other hand, the
tachyon amplitudes are non-vanishing only if a single tachyon has one
of the chiralities and the rest have opposite chirality.
Therefore the dual amplitude with the reduced variables (left side blob
in Fig.\ 2) vanishes
except when it is the three-point function ($k=2$).
This is precisely the case we have evaluated already
in Eq.\ \tacpol.
\par
As for the intermediate topological states, kinematics
dictates that it is of S-type for the dual amplitude with the
original variables corresponding to the right side blob in Fig.\ 2
$$
\eqalign{
{\bf p} & =
\biggl( -{N-k-1 \over \sqrt{2}}+{n \over \sqrt2(N-k-1)}\, , \cr
& \qquad\qquad
-i\left(-\sqrt2+ {N-k-1 \over \sqrt{2}}+{n \over \sqrt2(N-k-1)}
\right) \biggr)
}\eqn\kninttwomom
$$
for the level $n$ topological state.
As we have already explained, the intermediate state $-{\bf p}-{\bf Q}$
for the right side blob is of A-type.
The three-point function with the A-type topological state ($ k=2$)
is nothing but the OPE coefficient \tacope\ that we have seen
non-vanishing.
Four- and more- point functions with the A-type topological state
$ (k\ge 3)$  are more difficult to compute.
The topological state of level $ n$  consists of a linear combination
of monomials of derivatives of ${\bf X}$  multiplied by
a vertex operator $ {\rm e}^{i{\bf p} \cdot {\bf X}}$.
Both the number of $ \partial$  and the number of $ \bar \partial$
should be $ n$  for each monomial.
The two momentum $ {\bf p}$  is given by
$ ((r-t)/\sqrt2, -i(-2-r-t)/\sqrt2)$  for the $ (r,t)$
topological state of type A.
If we do not specify the coefficients of the monomials, we obtain
an operator containing the $ (r,t)$  topological state together with
a certain amount of null states.
We have taken such an operator as a substitute for the $ (r,t)$
topological state of A-type at the level $ n=rt$, and have explicitly
evaluated the dual amplitude with the
topological state for the case of four-point function.
We have found it to vanish.
This amplitude arises as the left side blob in Fig.\ 2 contributing
to the pole of level $ n=rt$  in the case of $ k=N-r-1=3$.
We conjecture in general that the A-type topological state
gives vanishing dual amplitude except for the three-point function
($ k=2$).
Only in the three-point dual amplitude ($ k=2$),
we can simply regard the factor for the blob of particles pinched
together (left side blob in Fig.\ 2)
as the coefficient of the OPE rather than the dual amplitude.
We should emphasize that this decoupling of the topological states
of the A-type is very
crucial in explaining the simple singularity structure of
the $N$-tachyon scattering amplitudes.
Let us note that the decoupling of these states has been
shown by Di Francesco and Kutasov in a recent
preprint\rlap.\refmark{\dfkulong}
This property is presumably related to Seiberg's finding that only
the S-type is physical.
These decoupling properties of both tachyon and
topological states originate partly from a peculiarity of the
two-dimensional kinematics (one dimension from the matter $X$ zero mode
and the other from the Liouville zero mode), but they also seem to be
a manifestation of the large symmetry
characteristic to the $c=1$ matter coupled to quantum gravity.
Hence they are worth studying further.
\par
Other possibilities are short-distance singularities from
the pinching of $ k$  tachyons all with positive
chirality, say $ {\bf p}_2, \cdots, {\bf p}_{k+1}$.
Actually the short-distance singularities due to the pinching
of tachyons all with positive chirality can be regarded as the same
short-distance singularities as the pinching of the other tachyons
instead.
Namely these limits are equivalent to configurations in which
the tachyons ${\bf p}_1$  and ${\bf p}_{k+2}, \cdots, {\bf p}_N$
are pinched rather than
${\bf p}_2, \cdots, {\bf p}_{k+1}$  by the $SL(2, {\bf C})$  invariance.
Since one of the other tachyons has the opposite chirality,
the present case is actually the same situation as the previously
analyzed case: pinching together the single negative chirality tachyon
with the opposite chirality tachyons.
Such configurations have already been discussed above and we need not
to consider them.
\par
These observations explain why there are only singularities in the
individual $p_j$, and none in any combinations of momenta, although the
factorization of the $N$-tachyon amplitudes is valid through the OPE as
we have seen.
\par
Let us finally discuss the case of $ s=$   positive integer.
The amplitudes with $ s=$  positive integer can be obtained
from the $ s = 0$  case as follows: we consider the $ N+s$
tachyon scattering amplitude and take a limit of vanishing
momenta for $ s$  tachyons and multiply by $ (\mu/\pi)^s$.
There is one subtlety: at the vanishing
momenta, the chirality is ill-defined\rlap.\refmark{\dfku) - \sataco}
We define the vanishing momenta taking the
limit from the positive chirality tachyon.
In the limit, we obtain an $s$-th power of a singular factor
$\mu \Delta(0)$, which should be replaced by the correct (renormalized)
cosmological constant $\mu_r$.
This procedure gives the insertion of the correct cosmological term
operator  $\phi \, {\rm e}^{\alpha\phi}$, as we explained earlier.
In this way we find that the short-distance singularities
of the amplitudes with non-vanishing $ s$  can be obtained correctly
once the short-distance singularities in the $ s = 0$  amplitude are
correctly obtained.
Using the previous argument, we find that the only non-vanishing
short-distance singularities come from the OPE
of two vertex operators for tachyons.
Short-distance singularities from one or more naive cosmological term
operators approaching the tachyon vertex operators give a
vanishing value for the residue.
\par
\vskip 5mm
\ack
One of the authors (NS) thanks Y. Kitazawa and D. Gross
for a discussion of the Liouville theory.
We would like to thank Patrick Crehan for a careful reading
of the manuscript.
This work is supported in part by Grant-in-Aid for Scientific Research
from the Ministry of Education, Science and Culture (No.01541237).
\vskip 5mm
\refout
\vskip 5mm
\FIG\a{The factorization of the $N$-tachyon amplitude by the OPE
       of the operators 1 and 2. The signs $+$  and $-$  denote
       the chirality of the tachyons.}
\FIG\b{The factorization of the $N$-tachyon amplitude by the OPE
       of the operators $1, \cdots, k$. The signs $+$  and $-$
       denote the chirality of the tachyons.}
\figout
\end